\documentclass[prd,preprint,showpacs,preprintnumbers,amsmath,amssymb,eqsecnum]{revtex4-1}

\usepackage[dvipdfmx]{graphicx}
\usepackage{bm}
\usepackage{color}
\usepackage{hhline}

\def\t#1{\tilde#1}
\def\b#1{\bar#1}

\begin{document}

\preprint{RUP-15-24}

\title{
High efficiency of collisional Penrose process requires heavy particle production
}

\author{$^{1}$Kota Ogasawara}
\email{k.ogasawara@rikkyo.ac.jp}
\author{$^{1}$Tomohiro Harada}
\email{harada@rikkyo.ac.jp}
\author{$^{2}$Umpei Miyamoto}
\email{umpei@akita-pu.ac.jp}

\date{\today}

\affiliation{$^{1}$Department of Physics, Rikkyo University, Toshima, Tokyo 171-8501, Japan}
\affiliation{$^{2}$RECCS, Akita Prefectural University, Akita 015-0055, Japan}

\begin{abstract}

The center-of-mass energy of two particles can become arbitrarily large
 if they collide near the event horizon of an extremal Kerr black hole,
 which is called the Ba$\rm \t n$ados-Silk-West (BSW) effect. 
We consider such a high-energy collision of two particles 
which started from infinity and follow geodesics in the equatorial plane
 and investigate the energy extraction from such a high-energy
 particle collision and the production of particles in the equatorial plane.
We analytically 
 show that, on the one hand, if the produced particles are 
 as massive as the colliding
 particles, the energy-extraction efficiency is bounded by
 $2.19$ approximately. On the other hand, if a very massive particle is produced as a result of the high-energy collision, which has negative energy and
 necessarily falls into the black hole, the upper limit of the 
 energy-extraction efficiency is increased to $(2+\sqrt{3})^2 \simeq 13.9$. 
Thus, higher efficiency of the energy extraction, which is 
typically as large as 10, 
provides strong evidence for the production of a heavy particle.

\end{abstract}


\maketitle
\newpage


\section{Introduction}

Ba$\rm \t n$ados, Silk, and West (BSW) pointed out that the
center-of-mass (CM) energy of two colliding particles can be arbitrarily
large, if the collision occurs near the event horizon of an extremal Kerr
black hole and the angular momentum of either of the colliding particles is
finetuned to the critical value~\cite{BSW}. 
This is now called the BSW effect. See Harada and
Kimura~\cite{Harada:2014vka} 
for a brief review and references therein for further details.
Particle collision with high CM energy had already been noticed by
Piran, Shaham, and Katz in the study of energy extraction from
collisional events in the ergoregion, which is called collisional
Penrose process~\cite{PiranShahamKatz,PiranShaham}. This process is
typically as follows. We consider the reaction of particles 1 and 2 into
particles 3 and 4 in the ergoregion, where particle 3 escapes to the
infinity after the collision, while particle 4 falls into the black hole
possibly with negative energy due to the existence of the ergosphere. 
If one defines an energy-extraction efficiency as
\begin{eqnarray}
\eta:=\frac{ {\rm energy~of~the~escaping~particle} }{ {\rm total~energy~of~the~injected~particles} }=\frac{E_3}{E_1+E_2},
\end{eqnarray}
the energy extraction ($\eta>1$) from the black hole is possible provided $E_4<0$.

The high-CM-energy collision can produce a very massive and/or energetic
particle. This means that Kerr black holes act as natural particle
accelerators, which can accelerate even neutral particles. Recently, the
interplay between such particle acceleration and energy extraction
have been intensively investigated~\cite{HNM,Schnittman,Berti:2014lva,Piran}. In particular, Schnittman \cite{Schnittman} numerically showed that the upper limit of energy-extraction efficiency in the collisional Penrose process can reach about 13.9. 
Berti, Brito, and Cardoso confirmed the result of
Ref.~\cite{Schnittman}, and also showed that an arbitrarily high efficiency is
possible by more general processes. Namely, they considered a head-on
collision of two subcritical particles, which is called 
super-Penrose process. This process needs an outgoing particle 
which must be generated in the ergoregion by some preceding 
process~\cite{Berti:2014lva,Piran}.

In this paper, we study particle collision near the horizon of an
extremal Kerr black hole and the resultant energy extraction. 
We present an analytic formulation to investigate collisional Penrose
process under the assumption that particles follow geodesics in the
equatorial plane and two particles collide near the horizon to produce
two particles. We find that if the
produced particles are as massive as the colliding particles, 
the energy-extraction efficiency is bounded by $2.19$ approximately. 
However, if a very massive particle is allowed to be produced, which
has negative energy and necessarily falls into the black hole,
the upper limit is increased to $(2+\sqrt{3})^2 \simeq 13.9$, confirming the numerical result in Ref.~\cite{Schnittman}.

The organization of this paper is as follows. In Sec.~\ref{Setup}, we
prepare for the analysis of collisional Penrose process, reviewing
geodesic motions and near-horizon collision in the Kerr black hole. In
Sec.~\ref{upper limit efficiency}, we investigate the upper limits of
the energy of escaping particle and of the energy-extraction efficiency
in the case of produced particles as massive as the colliding particles. 
In Sec.~\ref{sec.CM}, we will see that how the upper limits will be 
significantly increased if we take the production of a very massive
particles due to the BSW effect into account. 
Section \ref{Conclusion} is devoted to conclusion. 
We adopt the geometrized unit in which $c=G=1$.


\section{Preliminaries}
\label{Setup}


\subsection{Geodesics in the Kerr black hole}


The spacetime metric of the Kerr black hole is given by
\begin{eqnarray}
&& g_{\mu\nu} dx^\mu dx^\nu =
-\left(1-\frac{2Mr}{\rho^2}\right)dt^2-
\frac{4Mar\sin^2\theta}{\rho^2}dtd\varphi
+\frac{\rho^2}{\Delta}dr^2
+\rho^2 d\theta^2
\nonumber\\
&& \hspace{6cm} +\left(r^2+a^2+\frac{2Ma^2r\sin^2\theta}{\rho^2}\right)\sin^2\theta d\varphi^2,
\end{eqnarray}
where $\rho^2(r,\theta) := r^2+a^2\cos^2\theta$,
$\Delta(r):=r^2-2Mr+a^2$ and $M$ and $a \; (0 \leq a \leq M)$ are the mass and spin parameters,  respectively. $\Delta(r)$ vanishes at $r_\pm:=M\pm\sqrt{M^2-a^2}$, and $r=r_+$ and $r=r_-$ correspond to the event horizon and Cauchy horizon, respectively.

This spacetime is stationary and axisymmetric with Killing vectors
$\partial_t$ and $\partial_\varphi$. The conserved energy $E$ and
angular momentum $L$ of a particle with the four-momentum $p^\mu$ are given
by $E=-g_{\mu\nu}(\partial_t)^\mu p^\nu=-g_{t\mu}p^\mu$ and $L=g_{\mu\nu}(\partial_\varphi)^\mu p^\nu=g_{\varphi\mu}p^\mu$,
respectively. The components of the four-momentum are given in terms of
these conserved charges as (e.g.~\cite{Harada:2010yv,Harada:2011xz})
\begin{eqnarray}
&&p^t=\frac{1}{\Delta}\left[\left(r^2+a^2+\frac{2Ma^2}{r}\right)E-\frac{2Ma}{r}L\right],
\;\;\;
p^\varphi=\frac{1}{\Delta}\left[\frac{2Ma}{r}E+\left(1-\frac{2M}{r}\right)L\right],
\label{p^phi}
\\
&&\frac{1}{2}(p^r)^2+V=0,
\;\;\;
V(r)=-\frac{Mm^2}{r}+\frac{L^2-a^2(E^2-m^2)}{2r^2}-\frac{M(L-aE)^2}{r^3}-\frac{E^2-m^2}{2},
\nonumber \\
\label{V}
\end{eqnarray}
where $m \; (\geq 0)$ denotes the mass of the particle, and the motion
is assumed to be confined in the equatorial plane, where $\theta=\pi/2$.

The forward-in-time condition $p^t>0$ near the horizon $r\rightarrow r_+ + 0$ reduces to
\begin{eqnarray}
E-\Omega_HL\geq0,
\label{fit}
\end{eqnarray}
where $\Omega_H := a/(r^2_+ +a^2)$ is the angular velocity of the horizon.
We call a particle a critical particle if it has a critical angular
momentum $E/\Omega_H$, for which the equality in Eq.~\eqref{fit} holds. Accordingly, we call a particle with $L<E/\Omega_H $ ($L>E/\Omega_H$) a subcritical (supercritical) particle.

In the rest of this paper, we only consider the extremal black hole $a=M$. In this case, the forward-in-time condition for general position $r>r_+$ is written as
\begin{eqnarray}
\frac{1}{2}\left[\left(\frac{r}{M}\right)^3+\frac{r}{M}+2\right]E>\t L, 
\label{froward in time condition}
\end{eqnarray}
where $\t L:=L/M$ is a reduced angular momentum.

\begin{center}
\begin{figure}[bth]
\begin{tabular}{cc}
\includegraphics[width=0.46\textwidth]{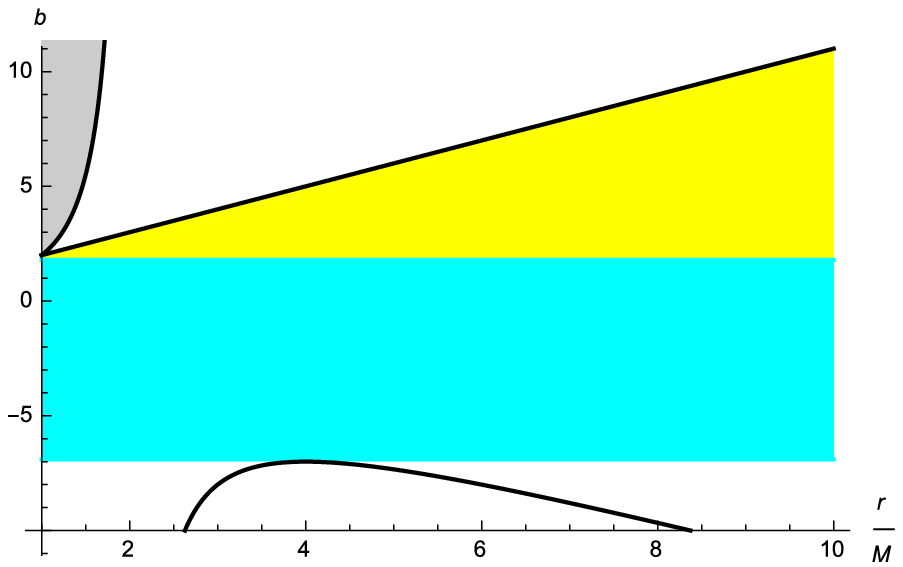}&
\includegraphics[width=0.46\textwidth]{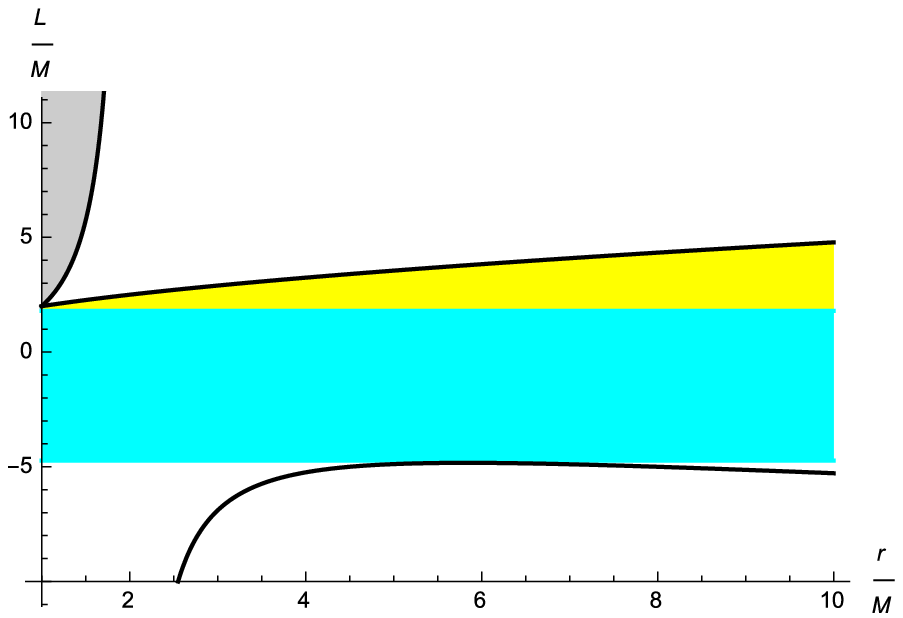} \\
(a) & (b) \\
\end{tabular}
\caption{(a) The radial turning points are plotted 
for a massless particle with $b=b_\pm
 (r)$. (b) The radial turning points are plotted for a massive particle
 with $\tilde{L}=\tilde{L}_\pm (r,E,m)$, where we set $E=m=1$. The negative
 energy particles are confined to the gray regions.}
\label{radial,turning,point}
\end{figure}
\end{center}

\subsection{Radial turning points of a geodesic particle}

Here, we are concerned with a particle that comes from or escapes to the
infinity, which requires the effective potential $V$ in Eq.~\eqref{V} is
non-positive for large $r$. This requires $E \geq m$.

For a massless particle  ($m=0$) , solving $V=0$ for the impact parameter $b:=L/E$, we obtain $b=b_\pm(r)$, where
\begin{eqnarray}
b_+(r) := r+M, \;\;\; b_-(r) := -\left(r+M+\frac{4M^2}{r-2M}\right).
\label{bpm}
\end{eqnarray}
This means that a particle of which impact parameter $ b= b_\pm(r)$ has
a turning point at $r$. The numerical plot of $b=b_\pm (r)$ is given in
Fig.~\ref{radial,turning,point}(a). As $r$ increases from $M$ to
infinity, $b_+(r)$ begins with $2M$ and monotonically increases to
infinity. As $r$ increases from $M$ to $2M$, $b_-(r)$ begins with $2M$
and monotonically increases to infinity. As $r$ increases from $2M$ to
infinity, $b_-(r)$ begins with negative infinity, monotonically
increases to a local maximum $-7M$ at $r=4M$ and monotonically decreases to negative infinity. Therefore, for
$2M<b<b_+(r_*)$, the particle can escape to the infinity irrespective of
the sign of the initial velocity, which is shown by the yellow region in
Fig.~\ref{radial,turning,point}(a) and where we denote the radial coordinate of
collision by $r_*$. On the other hand, for $M<r_{*}<4M$
and $-7M<b\leq 2M$, the
particle can escape to the infinity only if it moves initially
outwardly, which is shown by the blue region of Fig.~\ref{radial,turning,point}(a).

For a massive particle ($m>0$),
solving $V(r)=0$ for $\t L$, we obtain $\t L = \t L_\pm(r,E,m)$, where
\begin{eqnarray}
\t L_\pm(r,E,m) := \frac{-2M^2E\pm r(r-M)\sqrt{E^2-m^2+ 2Mm^2/r}}{M(r-2M)}.
\label{Lpm}
\end{eqnarray}
This means that a particle with $E$, $m$, and $\t L=\t L(r,E,m)$ has a
turning point at $r$. 
The numerical plot of $\t L = \t L_\pm(r,E,m) $ is given in
Fig.~\ref{radial,turning,point}(b). 
As $r$ increases from $M$ to infinity, $\t
L_+(r,E,m)$ begins with $2E$ and monotonically increases to
infinity. 
As $r$ increases from $M$ to $2M$,
$\t L_-(r,E,m)$ begins with $2E$ and monotonically increases to
infinity. As $r$ increases from $2M$ to infinity, $\t L_-(r,E,m)$ begins
with negative infinity, monotonically increases to a local maximum
$\t L_{\rm max}(E,m) \; (<0)$ at $r=r_{\rm max}$ and monotonically
decreases to negative infinity. Therefore, for $2E<\t L<\t
L_+(r_*,E,m)$, the particle can escape to infinity irrespective of the
sign of the initial velocity, which is shown by the yellow region of
Fig.~\ref{radial,turning,point}(b). On the other hand, for $M<r_{*}<r_{\rm
max}$ and $\t L_{\rm max}(E,m)<\t L\leq2E$, the particle can escape to
infinity only if it moves
initially outwardly, which is shown by the blue region of Fig.~\ref{radial,turning,point}(b).

\subsection{Particle collision on the horizon}
\label{Particle collision and reaction}

Let us consider the reaction of two colliding particles, named particles 1 and 2, to
two product particles, 3 and 4. The local conservation of four-momenta can be written as
\begin{eqnarray}
p^\mu_1+p^\mu_2=p^\mu_3+p^\mu_4.
\label{p^mu,conservation}
\end{eqnarray}
The $t$- and $\varphi$-components of Eq.~\eqref{p^mu,conservation} represent the conservations of energy and angular momentum
\begin{eqnarray}
E_1+E_2=E_3+E_4 
\quad \mbox{and}\quad  
\;\;\;
\t L_1+\t L_2=\t L_3+\t L_4,
\label{L,conservation}
\end{eqnarray}
respectively.
The $r$-component represents the conservation of radial momentum
\begin{eqnarray}
\sigma_1|p^r_1|+\sigma_2|p^r_2|=\sigma_3|p^r_3|+\sigma_4|p^r_4|,
\label{p^r,conservation}
\end{eqnarray}
where $\sigma_i = {\rm sgn}(p_i^r)$ for $i=1,2,3,4$.
Note that the mass and four-momentum of particle 4 can be written in terms of those of the other particles using the momentum conservations and identity $m^2_4=-p^\mu_4p_{4\mu}$. From Eq.~(\ref{V}), we obtain
\begin{eqnarray}
|p^r_i|=2E_i-\t L_i,
\end{eqnarray}
where we have used the forward-in-time condition to open the square root.

In the rest of this paper, we assume particle 1 to be critical ($\t L_1=2E_1$), particle 2 to be subcritical ($ \t L_2 < 2E_2 $), particle 3 to escape to infinity, and particle 4 to fall into the black hole with negative energy ($E_4<0$). Then, since particle 1 is critical, Eq. (\ref{p^r,conservation}) is written as
\begin{eqnarray}
\sigma_2(2E_2-\t L_2)
=
\begin{cases}
	\sigma_3(2E_2-\t L_2) & (\mbox{for $\sigma_3=\sigma_4$}) \\
	\sigma_3\left[2(2E_3-\t L_3)-(2E_2-\t L_2)\right] & (\mbox{for $\sigma_3=-\sigma_4$}) \\
\end{cases}.
\label{sigma_2(2E-L)}
\label{pr-cons}
\end{eqnarray}

When we choose $\sigma_2=-1 $, several situations are possible depending
on the values of $\sigma_3$ and $\sigma_4$. We will see, however, that only a few situations among them are interesting for our considerations. If $\sigma_3=\sigma_4=1$, from Eq.~\eqref{pr-cons}, we obtain $2E_2-\t L_2=0$, which contradicts our assumption. If $\sigma_3=\sigma_4=-1$, we obtain $2E_2-\t L_2=2E_3-\t L_3+2E_4-\t L_4$, which implies that particle 3 can be either critical ($\t E_3=2E_3$) or subcritical ($\t E_3<2E_3$). Nevertheless, only the critical case is interesting since a subcritical ingoing particle cannot escape to infinity. If $\sigma_3=- \sigma_4= 1$, we obtain $2E_3-\t L_3=0$ (particle 3 is critical). If $\sigma_3 = - \sigma_4= -1$, we obtain $2E_2-\t L_2=2E_3-\t L_3$, which implies particle 3 is subcritical, it is not interesting since a subcritical ingoing particle cannot escape to infinity again.

When we choose $\sigma_2=1$, only a few situations are interesting
again. 
If $\sigma_3=\sigma_4=1$, we obtain $2E_2-\t L_2=  (2E_3-\t L_3) + (2E_4-\t L_4) $, which implies that particle 3 can be either critical or subcritical.
Since $\sigma_3=1$ in the present case, particle 3 can escape to infinity even if it is subcritical and outgoing, provided $b_3$ or $\t L_3$ satisfy $b_{\rm max, 3}<b_3$ (massless case) or $\t L_{\rm max, 3}(E_3,m_3)<\t L_3$ (massive case). If $\sigma_3=\sigma_4=-1$, we obtain $2E_2-\t L_2=0$, which contradicts our assumption. If $\sigma_3=-\sigma_4=1$, we obtain $2E_2-\t L_2=2E_3-\t L_3$, which implies that particle 3 is subcritical. If $\sigma_3=-\sigma_4=-1$, we obtain $2E_3-\t L_3=0$.

From the above considerations, we see the following four situations are
interesting for the energy efficiency.
Case A: $\sigma_2=-1$ and particle 2 is subcritical. In this case, particles 1 and 2 come from infinity and particle 3 is critical at the horizon.
Case B: $\sigma_2=1$ and particle 2 is subcritical. In this case,
particle 2 must be created inside the ergoregion by some preceding process.
In the Appendix, we will see that the upper limit of the energy
extraction efficiency is in case B is the same as in case A.
Hence, we will focus on case A in Secs.~\ref{upper limit efficiency} and \ref{sec.CM}.

\subsection{Near-horizon and near-critical behaviors of a particle}

We parameterize the radial position of near-horizon collision $r_\ast$ as 
\begin{eqnarray}
	r_\ast = \frac{M}{1-\epsilon},
\;\;\;
	0<\epsilon \ll 1,
\label{r_ast}
\end{eqnarray}
and the near-critical angular momentum as
\begin{eqnarray}
\t L&=&2E(1+\delta),
\;\;\;
|\delta|\ll1 .
\end{eqnarray}
Then, we assume that $\delta$ can be expanded in powers of $\epsilon$ as
\begin{eqnarray}
\delta
=
\delta_{(1)}\epsilon+\delta_{(2)}\epsilon^2+ O(\epsilon^3).
\end{eqnarray}

Under the assumption that particle 3 escapes to the infinity, $\t
L_3\leq\t L_{+}(r_*,E_3,m_3)$ has to hold, which implies
\begin{eqnarray}
\delta\leq
\left(\frac{2E_3-\sqrt{E^2_3+m^2_3}}{2E_3}\right)\epsilon
+\left(\frac{4E_3\sqrt{E^2_3+m^2_3}-3E^2_3-2m^2_3}{2E_3\sqrt{E^2_3+m^2_3}}\right)\epsilon^2
+O(\epsilon^3).
\end{eqnarray}
The forward-in-time condition for the near-horizon and near-critical particle implies
\begin{eqnarray}
\delta<\epsilon+\frac{7}{4}\epsilon^2+O(\epsilon^3).
\end{eqnarray}
Therefore, the forward-in-time condition is always satisfied.

\subsection{Expansion of $|p^r_i|$ by $\epsilon$}

Let us consider the series expansion of the radial momentum in powers of
$\epsilon$ for each particle. 
Since we have assumed  particle 1 to be critical and particle 2 to be subcritical, $|p^r_1|$ and $|p^r_2|$ are expanded as
\begin{eqnarray}
|p^r_1|&=&
\sqrt{3E^2_1-m^2_1}\epsilon
-\frac{E^2_1}{\sqrt{3E^2_1-m^2_1}}\epsilon^2
+O(\epsilon^3),
\label{p1r}\\
|p^r_2|&=&
(2E_2-\tilde L_2)
-2(E_2-\tilde L_2)\epsilon
+\frac{(3E_2-\tilde L_2)(E_2-\tilde L_2)-m^2_2}{2(2E_2-\tilde L_2)}\epsilon^2
+O(\epsilon^3).
\label{p2r}
\end{eqnarray}
If particle 3 is near critical and particle 4 is subcritical, the expansions of $|p^r_3|$ and $|p^r_4|$ are given by
\begin{eqnarray}
|p^r_3|&=&
\sqrt{E^2_3\left[4(1-\delta_{(1)})^2-1\right]-m^2_3}\epsilon
-\frac{E^2_3\left[1-4(2\delta_{(1)}-\delta_{(2)})(1-\delta_{(1)})\right]}{\sqrt{E^2_3\left[4(1-\delta_{(1)})^2-1\right]-m^2_3}}\epsilon^2 + O(\epsilon^3), \nonumber \\
\label{p3r.critical}\\
|p^r_4|&=&
(2E_2-\tilde L_2)
-\left[2(E_2-\tilde L_2)+2E_3(1-\delta_{(1)})-2E_1\right]\epsilon
\nonumber\\&&~
+\left[
\frac{(2E_2-\tilde L_2)}{2}-2E_3(2\delta_{(1)}-\delta_{(2)})-\frac{(E_1+E_2-E_3)^2+m^2_4}{2(2E_2-\tilde L_2)}
\right]\epsilon^2
+O(\epsilon^3).
\label{p4rcritical}
\end{eqnarray}
If particle 3 is subcritical and particle 4 is near critical, the expansion of $|p^r_3|$ and $|p^r_4|$ are obtained by exchanging subscripts 3 and 4 in Eqs.~(\ref{p3r.critical}) and \eqref{p4rcritical}.
If both particles 3 and 4 are subcritical, the expansion of $|p^r_3|$ and $|p^r_4|$ are given by
\begin{eqnarray}
|p^r_3|&=&
(2E_3-\tilde L_3)
-2(E_3-\tilde L_3)\epsilon
+\frac{(3E_3-\tilde L_3)(E_3-\tilde L_3)-m^2_3}{2(2E_3-\tilde L_3)}\epsilon^2
+O(\epsilon^3),
\label{p3rsubcritical}\\
|p^r_4|&=&
(2E_2-\tilde L_2)-(2E_3-\tilde L_3)
+2(E_1-E_2+E_3+\t L_2-\t L_3)\epsilon
\nonumber\\
&&+
\frac{(E_1-E_2+E_3+\t L_2-\t L_3)(E_1+3E_2-3E_3-\t L_2+\t L_3)+m^2_4}{2[(2E_2-\tilde L_2)-(2E_3-\tilde L_3)]}\epsilon^2
+O(\epsilon^3).
\label{p4rsubcritical}
\end{eqnarray}

\section{Energy-extraction efficiency}

\subsection{Case for $m_{4}=O(\epsilon^{0})$}
\label{upper limit efficiency}
\label{The case(A)}

We focus on case A, where 
$\sigma_2=\sigma_4=-1$ and particle 3 is near critical. $\sigma_2=-1$
implies that we can assume that particles 1 and 2 come from the
infinity. The $O(\epsilon)$ and $O(\epsilon^2)$ terms of  radial momentum conservation Eq.~(\ref{p^r,conservation}) yield
\begin{eqnarray}
\sigma_1\sqrt{3E^2_1-m^2_1}
+2E_1-2E_3(1-\delta_{(1)})
=\sigma_3\sqrt{E^2_3 [4(1-\delta_{(1)})^2-1 ]-m^2_3},
\label{r.1.conservation.case(A)}
\end{eqnarray}
and
\begin{eqnarray}
\sigma_1\frac{E^2_1}{\sqrt{3E^2_1-m^2_1}}
+\frac{(3E_2-\tilde L_2)(E_2-\tilde L_2)-m^2_2}{2(2E_2-\tilde L_2)}
=
\sigma_3\frac{E^2_3 [1-4(2\delta_{(1)}-\delta_{(2)})(1-\delta_{(1)}) ]}{\sqrt{E^2_3 [4(1-\delta_{(1)})^2-1 ]-m^2_3}}
&&\nonumber\\ +
\frac{(2E_2-\tilde L_2)}{2}-2E_3(2\delta_{(1)}-\delta_{(2)})-\frac{(E_1+E_2-E_3)^2+m^2_4}{2(2E_2-\tilde L_2)},&&
\label{r.2.conservation.case(A)}
\end{eqnarray}
respectively.

When we choose $\sigma_1=1$, Eq.~(\ref{r.1.conservation.case(A)}) implies
\begin{eqnarray}
B_1-2E_3(1-\delta_{(1)})
=\sigma_3\sqrt{E^2_3 [ 4(1-\delta_{(1)})^2-1 ]-m^2_3},
\label{conserve1,+1,-1,-1}
\end{eqnarray}
where $B_1:=2E_1+\sqrt{3E^2_1-m^2_1} \; (>0)$.
Squaring the both sides of Eq.~\eqref{conserve1,+1,-1,-1}, we obtain
\begin{eqnarray}
1-\delta_{(1)}=\frac{B^2_1+E^2_3+m^2_3}{4B_1E_3},
\label{1-delta,+1,-1,-1}
\end{eqnarray}
which implies
\begin{eqnarray}
\delta_{(1),\rm max}-\delta_{(1)}=\frac{(B_1-\sqrt{E^2_3+m^2_3})^2}{4B_1E_3}\geq0.
\end{eqnarray}
Substituting Eq.~(\ref{1-delta,+1,-1,-1}) into the left-hand side of Eq.~(\ref{conserve1,+1,-1,-1}), we obtain
\begin{eqnarray}
B_1-\frac{E^2_3+m^2_3}{B_1}=2\sigma_3\sqrt{E^2_3 [4(\delta_{(1)}-1)^2-1 ]-m^2_3}.
\label{conserve1,B,+1,-1,-1}
\end{eqnarray}
This implies $E_3 \leq \tilde\lambda_0 := \sqrt{B^2_1-m^2_3}$ ($E_3 \geq \tilde\lambda_0$) for $\sigma_3=1$ ($\sigma_3=-1$).

If we choose $\sigma_3=-1$, we need $\delta_{(1)}\geq0$ since particle 3 has to be scattered by the potential barrier.
Supposing $\delta_{(1)}\geq0$ in Eq.~(\ref{1-delta,+1,-1,-1}), we have
\begin{eqnarray}
E^2_3-4B_1E_3+B^2_1+m^2_3\leq0.
\label{E2ineq}
\end{eqnarray}
Inequality~\eqref{E2ineq} is satisfied by 
\begin{eqnarray}
\t\lambda_-\leq E_3\leq\t\lambda_+,
\;\;\;
\t\lambda_\pm:=2B_1\pm\sqrt{3B^2_1-m^2_3},
\label{t,lambda_-,E,t,lambda_+}
\end{eqnarray}
where the discriminant $D$ of Eq.~\eqref{E2ineq} has to satisfy $D/4=3B^2_1-m^2_3\geq0$. Since particle 3 escapes to infinity, it is marginally bound or unbound ($E_3\geq m_3$). $m_3\leq\t\lambda_+$ has to be satisfied so that Eq.~(\ref{t,lambda_-,E,t,lambda_+}) and $E_3\geq m_3$ have an intersection.
The relation $m_3\leq\t\lambda_+$ is satisfied if $m_3\leq\sqrt{3}B_1$, which is equivalent to $D/4\geq0$.
Therefore, if $D/4\geq0$ is satisfied, Eq.~(\ref{t,lambda_-,E,t,lambda_+}) and $E_3\geq m_3$ always have an intersection.

$\t\lambda_0$ and $\t\lambda_+$ are the upper limits on $E_3$ for $\sigma_3=1$ and $\sigma_3=-1$, respectively.
Since $\t\lambda_+$ is larger than $\t\lambda_0$, we concentrate on the case of $\sigma_3=-1$.
The maximum of $\t\lambda_+$ is given by
\begin{eqnarray}
\t\lambda_{+,\rm max}=(2+\sqrt{3})^2E_1,
\end{eqnarray}
where we have assumed $m_1=m_3=0$.

Next, we consider the $O(\epsilon^2)$ terms in the radial momentum conservation. $E_3=\t\lambda_+$ can be realized when $\delta_{(1)}=0$.
Substituting $\delta_{(1)}=0$, $\sigma_1=1$, and $\sigma_3=-1$ into Eq.~(\ref{r.2.conservation.case(A)}), and then solving it for $m^2_4$, we obtain
\begin{eqnarray}
m^2_4&=&
-2 (2E_2-\t L_2 )\Big[
\frac{E^2_1}{\sqrt{3E^2_1-m^2_1}}
+\frac{\t\lambda^2_+}{\sqrt{3\t\lambda^2_+-m^2_3}}
+\frac{2\delta_{(2)}\t\lambda_+}{\sqrt{3\t\lambda^2_+-m^2_3}}\left(2\t\lambda_+-\sqrt{3\t\lambda^2_+-m^2_3}\right)
\Big]\nonumber\\
&& \hspace{8cm}
+(E^2_2+m^2_2)-(E_1+E_2-\t\lambda_+)^2.
\label{m^2_4,+1,-1,-1,-1}
\end{eqnarray}
We need $\delta>0$ for particle 3 to escape to the infinity. Since we choose $\delta_{(1)}=0$, we need $\delta_{(2)}\geq0$. This implies the first term on the right-hand side of Eq.~(\ref{m^2_4,+1,-1,-1,-1}) is negative.
The lower limit of $E_2$ is then given by
\begin{eqnarray}
E_2 \geq \t\kappa := \frac{1}{2}\left[(\t\lambda_+-E_1)-\frac{m^2_2}{(\t\lambda_+-E_1)}\right].
\label{E_2,min}
\end{eqnarray}
Since we assume that particle 2 comes from infinity, it must be
marginally bound or unbound ($E_2\geq m_2$). Therefore, we have to
compare $\t\kappa$ with $m_2$. If $\t\kappa\geq m_2$, i.e., 
$(\sqrt{2}-1)(\t\lambda_+-E_1)\geq m_2$, the lower limit of $E_2$ is $\t\kappa$. Thus, we find
\begin{eqnarray}
\eta\leq\frac{\t\lambda_+}{E_1+\t\kappa}
=\frac{2\t\lambda_+(\t\lambda_+-E_1)}{\t\lambda^2_+-E^2_1-m^2_2}=:f(m_2).
\end{eqnarray}
One can see that $f(m_2)$ defined above begins with $2\t\lambda_+/(\t\lambda_++E_1)$ and monotonically increases to
\begin{eqnarray}
\frac{(3+2\sqrt{2})\t\lambda_+}{(2+\sqrt{2})E_1+(1+\sqrt{2})\t\lambda_+}=:g(\t\lambda_+),
\end{eqnarray}
as $m_2$ increase from $0$ to $(\sqrt{2}-1)(\t\lambda_+-E_1)$.
Since $g(\t\lambda_+)$ is a monotonically increasing function of $\t\lambda_+$, the maximum of $g(\t\lambda_+)$ is given by
\begin{eqnarray}
g(\t\lambda_{+,\rm max})=
\frac{179+186\sqrt{2}+88\sqrt{3}+100\sqrt{6}}{383}
\simeq2.19.
\end{eqnarray}
If $\t\kappa\leq m_2$, i.e., $(\sqrt{2}-1)(\t\lambda_+-E_1)\leq m_2$, the lower limit of $E_2$ is $m_2$. Thus, we find
\begin{eqnarray}
\eta\leq\frac{\t\lambda_+}{E_1+m_2},
\end{eqnarray}
and easily notice that the right-hand side monotonically decreases as $m_2$ increase. For the above reason, the maximum of the right-hand side is about $2.19$.

In summary, the upper limit of the energy extraction efficiency is $\eta_{\rm max}\simeq2.19$, which is realized $\sigma_1=1$, $\sigma_2=\sigma_3=\sigma_4=-1$, $m_1=m_3=0$, and $\delta_{(1)}=0$.

Next, let us consider the case of $\sigma_1=-1$. Equation (\ref{r.1.conservation.case(A)}) implies
\begin{eqnarray}
A_1-2E_3(1-\delta_{(1)})
=\sigma_3\sqrt{E^2_3\Big[4(1-\delta_{(1)})^2-1\Big]-m^2_3},
\label{conserve-1,+1,-1,-1} 
\end{eqnarray}
where $A_1:=2E_1-\sqrt{3E^2_1-m^2_1} \; (>0)$.
Performing the following replacement, an argument similar to that in the
$\sigma_1=1$ case applies.
\begin{eqnarray*}
B_1&\rightarrow&A_1:=2E_1-\sqrt{3E^2_1-m^2_1},\\
\t\lambda_0&\rightarrow&\lambda_0:=\sqrt{A^2_1-m^2_3},\\
\t\lambda_\pm&\rightarrow&\lambda_\pm:=2A_1\pm\sqrt{3A^2_1-m^2_3},\\
\t\kappa&\rightarrow&\kappa:=\frac{1}{2}\left[(\lambda_+-E_1)-\frac{m^2_2}{(\lambda_+-E_1)}\right].
\end{eqnarray*}
However, the value of the upper limit is different. Since $\lambda_+$ is larger than $\lambda_0$, we concentrate on the case $\sigma_3=-1$.
The maximum of $\lambda_+$ is given by
\begin{eqnarray}
\lambda_{+, \rm max}=(2+\sqrt{3})(2-\sqrt{2})E_1,
\end{eqnarray}
where we have assumed $m_1=E_1$ and $m_3=0$.
We have seen that $E_3=\lambda_+$ can be realized $\delta_{(1)}=0$.
Substituting $\delta_{(1)}=0$ and $\sigma_1=\sigma_3=-1$ into Eq.~(\ref{r.2.conservation.case(A)}), and then solving it for $m^2_4$, we obtain
\begin{eqnarray}
m^2_4&=&
-2\left(2E_2-\t L_2\right) \Big[
-\frac{E^2_1}{\sqrt{3E^2_1-m^2_1}}
+\frac{\lambda^2_+}{\sqrt{3\lambda^2_+-m^2_3}}
+\frac{2\delta_{(2)}\lambda_+}{\sqrt{3\lambda^2_+-m^2_3}}\left(2\lambda_+-\sqrt{3\lambda^2_+-m^2_3}\right)
\Big]\nonumber\\
&& \hspace{8cm}
+(E^2_2+m^2_2)-(E_1+E_2-\lambda_+)^2.
\label{m^2_4,-1,-1,-1,-1}
\end{eqnarray}
Again, we need $\delta>0$ for particle 3 to escape to infinity. Since we choose $\delta_{(1)}=0$, we need $\delta_{(2)}\geq0$.
In this case, we can prove the first term on the right-hand side of Eq.~(\ref{m^2_4,-1,-1,-1,-1}) is negative (See Ref.~\cite[Appendix B]{HNM}).
The lower limit of $E_2$ is given by
\begin{eqnarray}
E_2\geq \kappa := \frac{1}{2}\left[(\lambda_+-E_1)-\frac{m^2_2}{(\lambda_+-E_1)}\right].
\end{eqnarray}
From the discussion similar to that in the case of $\sigma_1=1$, the upper limit of the energy extraction efficiency is also obtained as
\begin{eqnarray}
\eta_{\rm max}=\frac{2+\sqrt{2}+\sqrt{6}}{4}\simeq1.47,
\end{eqnarray}
which is realized $\sigma_1=\sigma_2=\sigma_3=\sigma_4=-1$, $E_1=m_1$, $m_3=0$, $\delta_{(1)}=0$, and $m_2/m_1\simeq0.491$.
This reproduces the result of Ref.~\cite{HNM}.

\subsection{Case for $m_{4}=O(\epsilon^{-1/2})$}
\label{sec.CM}
In general, the CM energy of particles 1 and 2 is given by
\begin{eqnarray}
E^2_{\rm cm}
=-(p^\mu_1+p^\mu_2)(p_{1\mu}+p_{2\mu}).
\end{eqnarray}
For example, in the original BSW process~\cite{BSW}, in which $\tilde{L}_1=2E_{1}$, $\tilde{L}_2<2E_{2}$ and $\sigma_1=\sigma_2 =-1$, the leading term of the CM energy is
\begin{eqnarray}
E^2_{\rm cm}
\simeq
\frac{2(2E_2-\t L_2)(2E_1-\sqrt{3E^2_1-m^2_1})}{\epsilon}.
\end{eqnarray}
In such a high-energy collision, the masses of the product particles (particles 3 and 4) can become large with the following restriction 
\begin{eqnarray}
m_3+m_4\leq E_{\rm cm}.
\end{eqnarray}
Since $E_{\rm cm}$ for the collision between a critical particle and a subcritical particle is proportional to $\epsilon^{-1/2}$ for both a rear-end collision and head-on collision, we assume 
\begin{eqnarray}
 m_4 = O(\epsilon^{-1/2}).
\end{eqnarray}
Then, we assume that particle 4 is very massive as
\begin{eqnarray}
m^2_4=\frac{\mu_4}{\epsilon}+\nu_4,
\label{heavy,m_4}
\end{eqnarray}
where $\mu_4 \; (>0)$ and $\nu_4$ are constants. 
In fact, we can write $m^2_4$ in terms of the quantities of the other particles, using the momentum conservation, as
\begin{eqnarray}
m^2_4 
=-(p^\mu_1+p^\mu_2-p^\mu_3)(p_{1\mu}+p_{2\mu}-p_{3\mu})
=
E^2_{\rm cm}+m^2_3+2(p^\mu_1+p^\mu_2)p_{3\mu} 
\end{eqnarray}
where we have assumed $\t L_3=2E_3(1+\delta)$. This implies
\begin{eqnarray}
\mu_4 = 2(2E_2-\t L_2) \Big[2E_1-2E_3(1-\delta_{(1)})+\sigma_1\sqrt{3E^2_1-m^2_1}-\sigma_3\sqrt{E^2_3 [4(1-\delta_{(1)})^2-1]-m^2_3} \Big].
\nonumber\\
\label{mu_4}
\end{eqnarray}
The expansion of radial momentum of particle 4 is given by
\begin{eqnarray}
|p^r_4|&=&
(2E_2-\t L_2)
-\left[2(E_2-\t L_2)+2E_3(1-\delta_{(1)})-2E_1+\frac{\mu_4}{2(2E_2-\t
  L_2)}\right]\epsilon \nonumber \\
&&+F_{4,\epsilon^2}\epsilon^2+O(\epsilon^3),
\end{eqnarray}
where 
\begin{eqnarray}
F_{4,\epsilon^2}&:=&
\frac{(2E_2-\tilde L_2)}{2}-2E_3(2\delta_{(1)}-\delta_{(2)})-\frac{(E_1+E_2-E_3)^2+\nu_4}{2(2E_2-\tilde L_2)}
\nonumber\\
&& 
+\frac{[E_1-E_2+\t L_2-E_3(1-\delta_{(1)})]\mu_4}{(2E_2-\tilde L_2)^2}
-\frac{\mu^2_4}{2(2E_2-\tilde L_2)^3}.
\end{eqnarray}
This implies that the $O(\epsilon)$ and $O(\epsilon^2)$ terms of the radial momentum conservation yield Eq.~(\ref{mu_4}) and
\begin{eqnarray}
\sigma_1\frac{E^2_1}{\sqrt{3E^2_1-m^2_1}}
+\frac{(3E_2-\tilde L_2)(E_2-\tilde L_2)-m^2_2}{2(2E_2-\tilde L_2)}-F_{4,\epsilon^2}
=
\sigma_3
\frac{E^2_3 [1+4(2\delta_{(1)}-\delta_{(2)})(\delta_{(1)}-1) ]}{\sqrt{E^2_3 [4(\delta_{(1)}-1)^2-1 ]-m^2_3}},
\nonumber\\
\label{p.2.momentum.BSW}
\end{eqnarray}
respectively.

If we define
\begin{eqnarray}
C:=2E_1+\sigma_1\sqrt{3E^2_1-m^2_1}-\frac{\mu_4}{2(2E_2-\t L_2)}>0,
\end{eqnarray}
we can discuss the upper limit of $E_3$ in the way similar to that in Sec.~\ref{The case(A)}.
As we have already seen,  we need $\sigma_3=-1$ and $\delta_{(1)}\geq0$ to obtain the maximum of $E_3$.
Setting $\delta_{(1)}\geq0$ in Eq.~(\ref{mu_4}), we have
\begin{eqnarray}
E^2_3-4CE_3+C^2+m^2_3\leq0.
\end{eqnarray}
This inequality is satisfied by
\begin{eqnarray}
\b\lambda_-\leq E_3\leq\b\lambda_+,
\;\;\;
\b\lambda_\pm:=2C\pm\sqrt{3C^2-m^2_3}.
\end{eqnarray}
The maximum value of $\b\lambda_+$ is given by
\begin{eqnarray}
\b\lambda_{+,\rm max}=(2+\sqrt{3})^2E_1-\frac{(2+\sqrt{3})\mu_4}{2(2E_2-\t L_2)},
\label{lambda+max}
\end{eqnarray}
where we have assume $m_1=m_3=0$ and $\sigma_1=1$.
$E_3=\b\lambda_+$ is realized when $\delta_{(1)}=0$.
Substituting $\delta_{(1)}=0$, $\sigma_1=1$, and $\sigma_3=-1$ into Eq.~(\ref{p.2.momentum.BSW}), and then solving it for $\nu_4$, we obtain
\begin{eqnarray}
\nu_4&=&
-2 (2E_2-\t L_2 ) \Big[
\frac{E^2_1}{\sqrt{3E^2_1-m^2_1}}
+\frac{\b\lambda^2_+}{\sqrt{3\b\lambda^2_+-m^2_3}}
+\frac{2\delta_{(2)}\b\lambda_+}{\sqrt{3\b\lambda^2_+-m^2_3}} \big( 2\b\lambda_+-\sqrt{3\b\lambda^2_+-m^2_3} \big)
\Big]\nonumber\\&&~
+(E^2_2+m^2_2)-(E_1+E_2-\b\lambda_+)^2+\frac{2(E_1-E_2+\t L_2-\b\lambda_+)\mu_4}{(2E_2-\t L_2)}-\frac{\mu^2_4}{(2E_2-\t L_2)^2}.
\label{nu_4=}
\end{eqnarray}
In Sec.~\ref{The case(A)}, we have used the above equation 
with $\mu_{4}=0$ and $\nu_{4}=m_{4}^{2}\ge 0$
to estimate the lower limit of $E_2$. 
However, in the present case, the sign of $\nu_4$ is not restricted 
to be positive if $\mu_{4}>0$. Hence, 
we use Eq.~(\ref{nu_4=}) not for the estimate of the lower
limit of $E_2$ but for the determination of $\nu_4$.
Therefore, the upper limit of the energy-extraction efficiency is given by
\begin{eqnarray}
\eta_{\rm max}=\frac{\b\lambda_{+,\rm max}}{E_1+E_2}
=\frac{(2+\sqrt{3})^2E_1-\frac{(2+\sqrt{3})\mu_4}{2(2E_2-\t L_2)}}{E_1+E_2}.
\end{eqnarray}
From the above equation,  we see that if $E_1E_2\gg\mu_4$, $\eta_{\max}$ is approximately given by $(2+\sqrt{3})^2E_1/(E_1+E_2)$.
Moreover, for $E_2/E_1 \ll 1$, we find that the escaping
massless particle has the energy that is nearly equal to 
$(2+\sqrt{3})^{2}\simeq 13.9$ times the total energy of 
the incoming massless particles.

However, $E_3$ cannot be exactly equal to $(2+\sqrt{3})^2E_1$, for the
following reason.
From Eq.~(\ref{lambda+max}), $E_3=(2+\sqrt{3})^2E_1$ is realized when $m_1=m_3=0$, $\delta_{(1)}=0$ and $\mu_4=0$.
However, $\mu_4=0$ implies $m_4^{2}=O(\epsilon^{0})$ and hence 
$\nu_4$ must be positive. Substituting
$\mu_4=0$ into Eq.~(\ref{nu_4=}) and requiring $\nu_4>0$, we obtain the
lower limit of $E_2$, which is given by Eq.~(\ref{E_2,min}). In this
case, we have already seen that the efficiency cannot reach
$(2+\sqrt{3})^2\simeq 13.9$ but only 2.19.

While there are other cases in which a very massive and/or energetic
particle is produced, one can conclude that a large efficiency is 
possible only in the above case.
In any other cases, $p^r_3$ and $p^r_4$ have $O(\epsilon^{1/2})$
terms. Since there is no half-integer order terms on the left-hand side
of Eq.~(\ref{p^r,conservation}), $O(\epsilon^{1/2})$ term in the sum of
$p^r_3$ and $p^r_4$ has to be zero. In fact, if one assumes particle 3
to be very energetic so that $E^2_3={\rho_3}/{\epsilon}+\phi_3$, where $\rho_3 \; (>0)$ and $\phi_3$ are constants, the $O(\epsilon^{1/2})$  terms in the radial momentum conservation yield
\begin{eqnarray}
\sigma_3\sqrt{\rho_3 [4(1-\delta_{(1)})^2-1 ]}-2(1-\delta_{(1)})\sqrt{\rho_3}=0.
\end{eqnarray}
There is no solution for $\rho_3$ under the assumption of $\rho_3>0$.
By similar arguments, one can conclude that there is no energy
extraction, expect for the case where particle 4 is very massive.

\section{Conclusion}
\label{Conclusion}

We have studied particle collision and energy-extraction efficiency,
where a critical particle (particle 1) and 
subcritical particle (particle 2) collide near the event horizon of an
extremal Kerr black hole and then two particles are produced, one of which 
escapes to infinity (particle 3) and another falls into the black hole
(particle 4).

There is an upper limit of the energy of particle 3, which is given by 
$E_{3,{\rm max}}=(2+\sqrt{3})^2E_1$.
This is realized in the situation where particles 1 and 3 are massless,
particle 3 is near critical, 
particle 1 is temporary outgoing and particles 2, 3 and 4
are temporarily ingoing.
The energy-extraction efficiency,
however, is bounded by $ 2.19$ approximately, under the assumption
that particle 4 has a mass of $O(\epsilon^{0})$, where $\epsilon$
parametrizes the distance from the collision point to the horizon.

Since the CM energy of the near-horizon collision can be arbitrarily
large, the collision can produce a heavy particle. From this
viewpoint, we have considered the case in which the mass of particle 4 
is of $O(\epsilon^{-1/2})$. 
In this case, we found that 
the upper limit of 
the efficiency can indeed reach $(2+\sqrt{3})^{2}\simeq 13.9$.
Thus, if the efficiency as large as 10 is observed for
a collisional Penrose process, this strongly suggests
the production of a heavy particle as a result of the collision of 
high CM energy.

Finally, let us give an example of collision with a high energy-extraction
efficiency. We assume that particles 1 and 2 are protons, 3 is a
photon, and 4 is some heavy particle and that 
the collision happens in the vicinity of the horizon with $\epsilon=10^{-8}$.
Using $m_{1}=m_{2} \sim $ 1 GeV, we have $E_{\rm
cm}\sim\sqrt{m_1m_2/\epsilon}\sim10^4$ GeV. If $m_4 \sim 10^3$ GeV,
we have $\mu_4\sim
m^2_4\epsilon\sim10^{-2}$ GeV$^2$ which is much smaller than $E_{1}E_{2}$. 
In this case, the efficiency is typically as large as 10.

{\bf Note added:} While completing the current paper, the authors found
that a paper~\cite{Leiderschneider:2015kwa} studying a similar problem
appeared in arXiv. It will be interesting to compare the result in
Ref.~\cite{Leiderschneider:2015kwa} with that in the current paper.


\acknowledgments
The authors would like to thank T.~Igata, M.~Kimura, T.~Kobayashi, K.~Nakao, M.~Patil 
and S.~Yokoyama.
This work was supported by JSPS KAKENHI Grant Numbers 26400282 (TH) and 
15K05086 (UM).


\appendix

\section{Collision between a critical particle and an outgoing
 subcritical particle}

Here we consider case B, where $\sigma_2=1$ and particle 2 is
subcritical. We can divide this case into the following three subcases: 
B1: $\sigma_2=\sigma_4=1$ ($\sigma_1$ and $\sigma_3$ can be either $1$ or $-1$).  Particle 3 is critical.
B2: $\sigma_2=\sigma_3=\sigma_4=1$.
Both particles 3 and 4 are subcritical, and $2E_2-\t L_2 = (2E_3-\t L_3)
+ (2E_4-\t L_4)$.
B3: $\sigma_2=\sigma_3=1$, $\sigma_4=-1$.
Particle 3 is subcritical, and $2E_2-\t L_2=2E_3-\t L_3$, which implies particle 4 to be critical.

For case B1, the $O(\epsilon)$ terms in the radial momentum conservation yield
\begin{eqnarray}
\sigma_1\sqrt{3E^2_1-m^2_1}-2E_1+2E_3(1-\delta_{(1)})
=\sigma_3\sqrt{E^2_3 [4(\delta_{(1)}-1)^2-1 ]-m^2_3}.
\end{eqnarray}
This is obtained from Eq.~(\ref{r.1.conservation.case(A)}) after replacing $\sigma_1$ and $\sigma_3$ with $-\sigma_1$ and $-\sigma_3$, respectively.
The equation obtained from the $O(\epsilon^2)$ terms of the radial
momentum conservation is also the same as
Eq.~(\ref{r.2.conservation.case(A)}) after the above
replacement. Therefore, $\eta_{\rm max}$ in this case is given by that
in Sec.~\ref{The case(A)} after changing the signs $\sigma_i$
($i=1,2,3,4$). Note that the change of $\sigma_3$ requires the argument
on the turning point, but one can finally see that 
this prescription is valid.

For case B2, the $O(\epsilon)$ terms in the radial momentum conservation equations yield
\begin{eqnarray}
\sigma_1\sqrt{3E^2_1-m^2_1}-2E_1=0.
\label{r.1.conservation.p3rsubcritical}
\end{eqnarray}
This equation has no solution for $E_1$ under the assumption $E_1>0$ and $m_1^2\geq0$. Thus, there is no energy extraction in this case.

For case B3, the $O(\epsilon)$ terms of the radial momentum conservation equations yield
\begin{eqnarray}
2E_1-\sigma_1\sqrt{3E^2_1-m^2_1}-2E_4(1-\delta_{(1)})=\sqrt{E^2_4 [4(\delta_{(1)}-1)^2-1 ]-m^2_4}.
\label{r.1.momentum.case(d)}
\end{eqnarray}
Equation (\ref{r.1.momentum.case(d)}) implies
\begin{eqnarray}
B_1-2E_4(1-\delta_{(1)})=\sqrt{E^2_4 [4(\delta_{(1)}-1)^2-1 ]-m^2_4}
\label{pr1,4near}
\end{eqnarray}
for $\sigma_1=-1$ and $-\sqrt{B^2_1-m^2_4}\leq E_4<0$.
From the forward-in-time condition and the argument on the turning point for a near-critical particle with negative energy, the restriction on $\delta_{(1)}$ is obtained as
\begin{eqnarray}
\delta_{(1)}\geq1-\frac{\sqrt{E^2_4+m^2_4}}{2E_4}.
\end{eqnarray}
Squaring the both sides of Eq.~(\ref{pr1,4near}) and then solving it for $E_4$, we obtain
\begin{eqnarray}
E_4=2(1-\delta_{(1)})B_1+\sqrt{B^2_1 [4(\delta_{(1)}-1)^2-1 ]-m^2_4}.
\end{eqnarray}
This implies the lower limit of $E_4$ is given by
\begin{eqnarray}
E_{4,\rm min}=-(2+\sqrt{3})E_1,
\end{eqnarray}
where $m_1=m_4=0$ and $\delta_{(1)}=3/2$ hold.
The $O(\epsilon^2)$ terms of the radial momentum conservation equation yield
\begin{eqnarray}
-\sigma_1\frac{E^2_1}{\sqrt{3E^2_1-m^2_1}}
+\frac{(3E_2-\tilde L_2)(E_2-\tilde L_2)-m^2_2}{2(2E_2-\tilde L_2)}
=
\frac{E^2_4 [1-4(2\delta_{(1)}-\delta_{(2)})(1-\delta_{(1)}) ]}{\sqrt{E^2_4 [4(\delta_{(1)}-1)^2-1 ]-m^2_4}}
\nonumber \\
+\frac{ 2E_2-\tilde L_2 }{2}-2E_4(2\delta_{(1)}-\delta_{(2)})-\frac{(E_1+E_2-E_4)^2+m^2_3}{2(2E_2-\tilde L_2)} .
\label{B3O2}
\end{eqnarray}
The lower limit of $E_4$ can be realized only for 
$m_1=m_4=0$ and $\delta_{(1)}=3/2$.
Because the denominator of the first term on the right-hand side of Eq.~(\ref{B3O2}) becomes zero when $m_4=0$ and $\delta_{(1)}=3/2$, we need $1-4(2\delta_{(1)}-\delta_{(2)})(1-\delta_{(1)})=0$, which is possible when $\delta_{(2)}=7/2$.
Therefore, the upper limit of the energy-extraction efficiency in this case is given by
\begin{eqnarray}
\eta_{\rm
 max}=1-\frac{E_{4,\min}}{E_1+E_2}=\frac{(3+\sqrt{3})E_1+E_2}{E_1+E_2}
<3+\sqrt{3},
\end{eqnarray}
where we have assumed $m_1=m_4=0$, $\delta_{(1)}=3/2$, and $\delta_{(2)}=7/2$.

If we assume $m_{4}=O(\epsilon^{-1/2})$, the energy-extraction
efficiency can be larger as in case A.
For case B1, the result is the same as for case A after replacing $\sigma_i$ with
$-\sigma_i$ ($i=1,2,3,4$). Therefore, we can obtain the upper limit 
$\simeq 13.9$, only if a very massive particle is produced to fall into the black hole.
For cases B2 and B3, the energy extraction is quite modest.
Namely, the efficient energy extraction is not realized even when 
an subcritical outgoing particles is considered, as long as 
its counterpart is critical.


\begin{thebibliography}{99}

\bibitem{BSW}
M.\ Ba${\rm \tilde n}$ados, J.\ Silk, and S.\ M.\ West, 
``Kerr Black Holes as Particle Accelerators to Arbitrarily High
	Energy'', 
Phys.\ Rev.\ Lett. \textbf{103} (2009) 111102 [arXiv:0909.0169 [hep-ph]].

\bibitem{Harada:2014vka}
  T.~Harada and M.~Kimura,
  ``Black holes as particle accelerators: a brief review,''
  Class.\ Quant.\ Grav.\  {\bf 31} (2014) 243001
  [arXiv:1409.7502 [gr-qc]].

\bibitem{PiranShahamKatz}
T.\ Piran, J.\ Shaham, and J.\ Katz, 
``High efficiency of the Penrose mechanism for particle collisions'', 
Astrophys. J. \textbf{196} (1975) L107.

\bibitem{PiranShaham}
T.\ Piran and J.\ Shaham, 
``Upper bounds on collisional Penrose processes near rotating black-hole
horizons'', 
Phys. Rev. D \textbf{16} (1977) 1615.

\bibitem{HNM}
T.\ Harada, H.\ Nemoto, and U.\ Miyamoto, 
``Upper limits of particle emission from high-energy collision and reaction
near a maximally rotating Kerr black hole'', 
Phys.\ Rev.\ D \textbf{86} (2012) 024027 [arXiv:1205.7088 [gr-qc]].

\bibitem{Schnittman}
J.\ D.\ Schnittman, 
``A revised upper limit to energy extraction from a Kerr black hole'', 
Phys.\ Rev.\ Lett.\ \textbf{113} (2014) 261102. [arXiv:1410.6446 [gr-qc]].

\bibitem{Berti:2014lva}
  E.~Berti, R.~Brito and V.~Cardoso,
  ``Ultrahigh-energy debris from the collisional Penrose process,''
  Phys.\ Rev.\ Lett.\  {\bf 114} (2015) 25,  251103
  [arXiv:1410.8534 [gr-qc]].

\bibitem{Piran}
E.\ Leiderschneider and T.\ Piran, 
``Super-Penrose collisions are inefficient - a Comment on: Black hole
fireworks: ultra-high-energy debris from super-Penrose collisions'', 
arXiv:1501.01984 [gr-qc].

\bibitem{Harada:2011xz}
  T.~Harada and M.~Kimura,
  ``Collision of two general geodesic particles around a Kerr black hole,''
  Phys.\ Rev.\ D {\bf 83} (2011) 084041
  [arXiv:1102.3316 [gr-qc]].

\bibitem{Harada:2010yv}
  T.~Harada and M.~Kimura,
  ``Collision of an innermost stable circular orbit particle around a Kerr black hole,''
  Phys.\ Rev.\ D {\bf 83} (2011) 024002
  [arXiv:1010.0962 [gr-qc]].

\bibitem{Leiderschneider:2015kwa} 
  E.~Leiderschneider and T.~Piran,
  ``On the maximal efficiency of the collisional Penrose process,''
  arXiv:1510.06764 [gr-qc].

\end{thebibliography}
\end{document}